# Alert of the Second Decision-maker: An Introduction to Human-AI Conflict

**He Wen**

Department of Civil and Environmental Engineering,

University of Alberta, Edmonton, AB T6G 2E1, Canada

**Abstract**
The collaboration between humans and artificial intelligence (AI) is a significant feature in this digital age. However, humans and AI may have observation, interpretation, and action conflicts when working synchronously. This phenomenon is often masked by faults and, unfortunately, overlooked. This paper systematically introduces the human-AI conflict concept, causes, measurement methods, and risk assessment. The results highlight that there is a potential second decision-maker besides the human, which is the AI; the human-AI conflict is a unique and emerging risk in digitalized process systems; and this is an interdisciplinary field that needs to be distinguished from traditional fault and failure analysis; the conflict risk is significant and cannot be ignored.

## 1. Introduction

Automation, digitization, and artificial intelligence (AI) have become the trends in the development of industrial history (Pistikopoulos et al., 2021). Significantly, the applications of AI in process industries have seen geometric growth (Lee et al., 2019). Digitization is profoundly changing the relationship between humans and intelligent machines or systems (Khan et al., 2021), which is referred to as AI here. When humans or AI work alone, they do not affect each other; but once they work simultaneously, the difference in cognition, decision-making, and execution between the two emerges, which may end up creating a conflict.

This conflict here is called human-AI conflict. As different people have various views of AI (Wang, 2008), the conflict in this study is more likely human-machine conflict. Yet, it fails to highlight the cognitive capabilities and situational awareness of intelligent machines and automated systems. Therefore, it should be called human-AI conflict for better understanding. It is a new type of risk brought about by system digitization and automation, and some accidents due to human-AI conflict have already occurred.

For example, self-driving cars may brake unexpectedly when wrongly detecting front obstacles, and this phenomenon is "phantom braking" (Moscoso Paredes et al., 2021). Similarly, the phenomenon, known as "automation surprise" (Boer & Dekker, 2017; Dehais et al., 2015; Sarter et al., 1997) or "mode confusion" (Bredereke & Lankenau, 2005; Leveson et al., 1997; Rushby, 2002), is also visible in the aviation industry. The Boeing 737 Max crashes are some of the notable examples in this regard (The House Committee on Transportation and Infrastructure, 2020). Briefly, it is because the senor fault activated the maneuvering characteristics augmentation system (MCAS) to push the aircraft into a dive, and it prevented the pilot from pulling up manually.

Human-AI conflict is also possible between the human operator and the controller for process control systems. As the process industry involves significant hazard sources, the conflict risk cannot be underestimated and should be highly prioritized. This paper tries to give the readers a systematic introduction to the human-AI conflict. The novelty of this paper is

i. Summary of the conflict causes and their formulation from safety and security perspectives (i.e., sensor fault, cyberattack, human error, and sabotage).
ii. Defining the variables of observation conflict, interpretation conflict, and action conflict.
iii. Proposing the distance between humans and AI to measure the conflict.
iv. Generalizing the assessment method of conflict risk.

The rest of this paper is organized as follows:

- Section 2 answers the general questions: 1) What is human-AI conflict? 2) Who are the participants of a conflict? 3) At what stage of industrial development does human-AI conflict occur? 4) What are the causes of human-AI conflict?
- Section 3 demonstrates how to formulate and measure the conflict by introducing new variables and extensions in multiple input situations.
- Section 4 illustrates the formulation of conflict probability, conflict severity, and conflict risk.
- Section 5 are discussion of current research problems and future directions, and some remarkable findings in the study of human-AI conflict.

## 2. Human-AI conflict
### 2.1. Definition

Human-AI conflict is the difference in the observation, interpretation, or action of one or more variables by different participants (the human operator and AI) (Wen, Amin, et al., 2022). A typical conflict evolution sequence is that observation difference leads to interpretation difference, which ultimately leads to action difference (Fig. 1). It is also possible that in the absence of observation conflict, interpretation conflict can occur, which then leads to action conflict (Wen, Amin, et al., 2023; Wen, Khan, et al., 2022, 2023). Or even a direct action conflict between human and AI occurs, supposing one of the two provokes the conflict. All the differences will be reflected in the observations at the next time-step; hence, this study initiates the discussion from observation conflict.

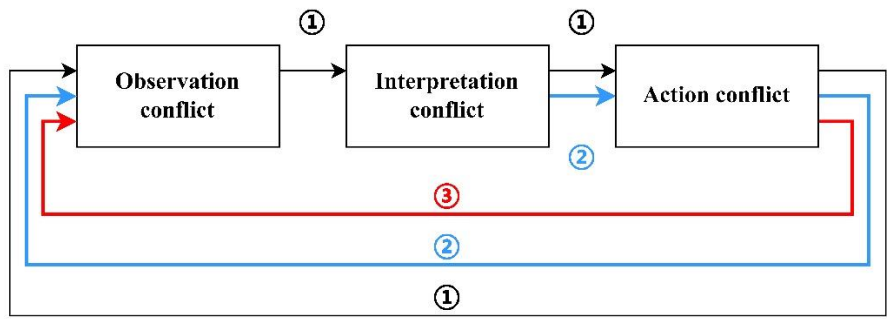

Fig. 1: Three paths of conflict evolution.

Usually, there will be no conflict when humans and AI work independently and separately. However, when they work together, there is a probability of a conflict occurring. Not all machines can conflict with humans; that must be those anthropomorphic machines, for example, the automated systems, which have perception ability, logic processing ability, and execution ability. A control system with a sensor, a controller, and an actuator, already has the basic units of AI. Therefore, we collectively refer to the automated systems of different intelligence stages as AI.

In the future, automated machines may completely replace humans, even without human supervision, and full automation will not lead to human-AI conflict. However, in all the other stages before full automation (Vagia et al., 2016), human-AI conflict may occur as long as humans and AI collaborate. It is even foreseeable that the more intelligent the AI, the more likely it is to compete with humans for priority.

### 2.2. Conflict cause

Fundamentally, human-AI conflict can arise for a variety of reasons. Common causes include sensor faults, cyberattacks, human errors, and sabotage (Fig. 2). Sensor faults and human errors can be categorized as safety areas, while cyberattacks and sabotage belong to the security field. This study does not consider conflicts between the operator and the sabotager because the topic is human-AI conflict, not human-human conflict.

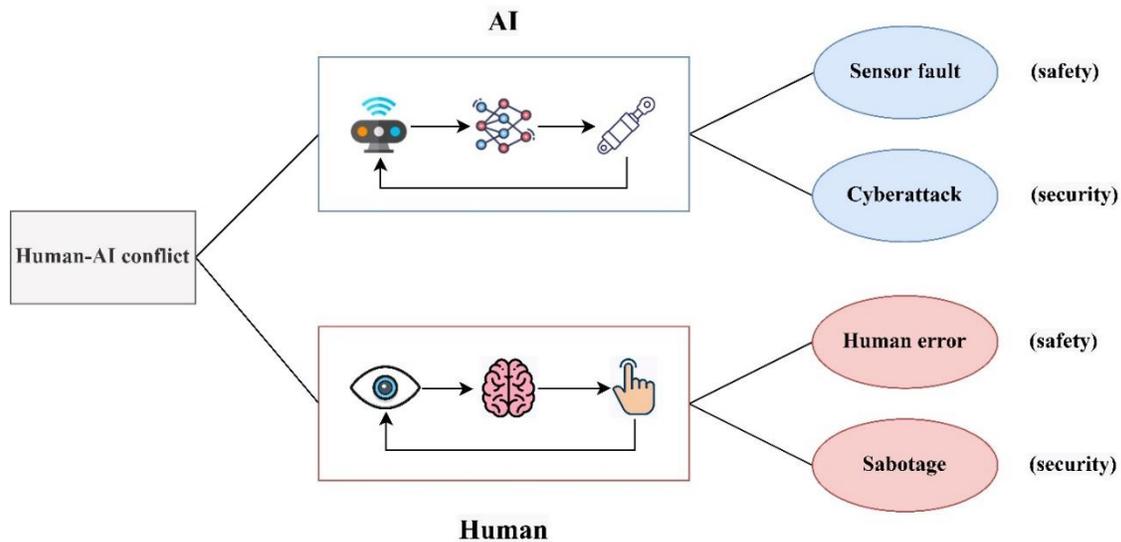

Fig. 2: Causes of human-AI conflict from safety and security perspectives (Wen, Khan, et al., 2023).

Comparatively, sensor faults are more common and traditional, such as open circuit, short circuit, stuck, bias, cyclic, and drift (Yung & Clarke, 1989). Various situations may trigger different faults, such as power off, electric short, aging, deterioration, contamination, internal mechanical change, or even environmental impact.

Recently, cyber threats have demonstrated a source of conflict. Frequent cyberattacks on the automated system include denial of service (DoS), false data injection (FDI) on sensor or actuator, setpoint modification, and time delay. Furthermore, some advanced attacks may destroy the logic solver's or AI's model and algorithm, such as adversarial attacks, backdoor attacks, data pollution attacks, etc.

In the traditional safety area, human error can proactively provoke conflicts. As the AI or the automated control system participates in most operations, the human operator only works as the supervisor to monitor the operations. Usually, no action from the operator is a kind of corrective performance. The error action may appear at start-up, shutdown, setting initial parameters, and maintenance.

Malicious sabotage can also create conflicts with the AI or the control systems. Insiders or external sabotagers can manipulate many types of faults. Similarly, the actions may include forced stop, physical destruction, parameter change maliciously, etc. The sabotage methods are more unpredictable than human error.

## 3. Conflict measurement

### 3.1. Variable of observation difference

To measure the conflict, this study introduces a variable of observation difference (VOD), which is the difference in observation of process value from different observers (Wen, Amin, et al., 2022). Suppose sensor observation of AI or automated control system as $x_A(t)$, human observation as $x_H(t)$.

$$VOD = x_A(t) - x_H(t) \tag{1}$$

Suppose sensor fault as $f_S(t)$, cyberattack as $f_C(t)$, human error as $f_H(t)$, sabotage as $f_I(t)$, the normal status of human observation and AI observation as $x_N(t)$. Faults can represent the final manifestation of all abnormal situations (Wen, Amin, et al., 2023).

$$x_A(t) = x_N(t) + f_S(t) + f_C(t) \tag{2}$$

$$x_H(t) = x_N(t) + f_H(t) + f_I(t) \tag{3}$$

Substitute Eq. (2) and Eq. (3) into (1)

$$VOD = f_S(t) + f_C(t) - f_H(t) - f_I(t) \tag{4}$$

This means that observation conflict is just the abnormal additions by the conflict causes. Some common types of conflict causes are formulated in Table 1, which also gives the generalized mathematical expressions on sensor observation. Based on these equations of sensor observation, the VODs can be derived correspondingly, which are shown in Fig. 3.

Table 1. Conflict causes and mathematical expressions on sensor observation.

| Sensor fault | Cyberattack | Human error | Sabotage | Mathematical expression |
|---|---|---|---|---|
| Open circuit | FDI | Error to stop | Force to stop | $x_A(t) = 0$ |
| Short circuit | FDI | Error in connecting wires | Malicious wire connection | $x_A(t) = \infty$ |
| Stuck | FDI | Wrong input | Physical destruction | $x_A(t) = x_A(t_0)$ |
| Bias | FDI, Setpoint modification | Mistake input | Malicious change of parameters | $x_A(t) = x_N(t) \pm \Delta$ |
| Cyclic | FDI, DoS | No input when required | Intentional inaction, cut-off network, malicious contamination | $x_A(t) = x_N(t) + e$ |
| Drift | FDI | Mistake input | Physical destruction | $x_A(t) = x_N(t) + e(t)$ |
| - | Time delay | - | Physical destruction | $x_A(t) = \begin{cases} x_N(t) + e(t), & t \in [0, \tau] \\ x_N(t), & t \in (\tau, \infty) \end{cases}$ |

Where $t_0$ is the time to occur an abnormal situation, $\Delta$ is a constant, $e$ is a random error, $e(t)$ is a changing error, $\tau$ is the delayed time.

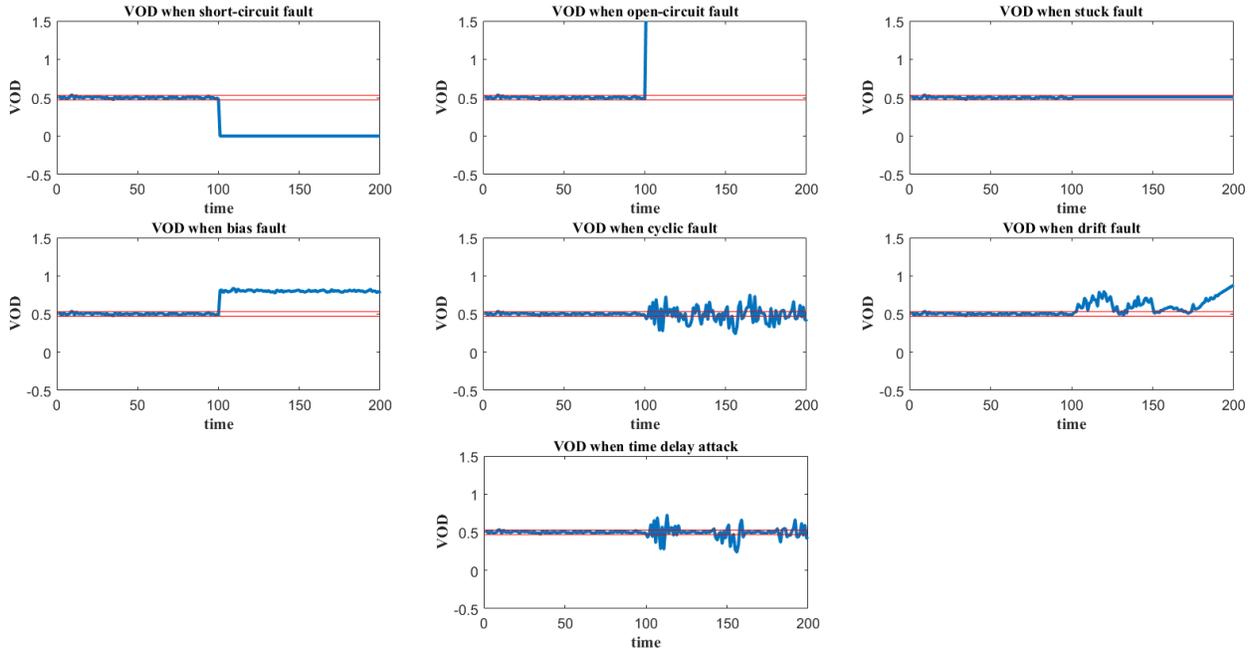

Fig. 3: Schematic diagrams of VODs (assume the abnormal situation occurs at 100s).

The above illustration is based on one process variable. In one-dimensional space, the observation difference between human and AI is the Manhattan distance between the two.

$$d_{VOD} = |x_{Ai}(t) - x_{Hi}(t)| \tag{5}$$

Where $d$ is distance.

For multiple input multiple output, the observation conflict is

$$VOD = X_A(t) - X_H(t) = \begin{bmatrix} x_{A1}(t) \\ x_{A2}(t) \\ \ldots \\ x_{An}(t) \end{bmatrix} - \begin{bmatrix} x_{H1}(t) \\ x_{H2}(t) \\ \ldots \\ x_{Hn}(t) \end{bmatrix} \tag{6}$$

Where $X_A(t)$ is sensor observation matrix, $X_H(t)$ is human observation matrix; the system has $n$ observable process variables.

Compared with the Manhattan distance in one-dimensional space, when extended to multivariate, Euclidean distance will be more applicable.

$$d_{VOD} = \sqrt{\sum_{i=1}^{n}(x_{Ai}(t) - x_{Hi}(t))^2} \tag{7}$$

### 3.2. Variable of interpretation difference

As the human cognition process is unknown, it is quite difficult to measure the interpretation conflict. Nevertheless, we still see many conflict situations, such as automation surprise and mode confusion. All the zone between confirmed correct and convinced wrong is interpretation conflict.

For AI, most inputs/observations are converted to a matrix for calculation, for example, the task of image recognition. Specifically, AI performs matrix calculations with neural networks to get a score vector and transfers the score to the interpretation probability vector $P(y_A)$ with Softmax function, and then conclude the interpretation classification result vector $y_A$ by cross-entropy function, where $y_A$ is a one-hot vector.

For humans, assuming the above process is also true, human observations are also converted into a matrix; hence, the general form of VOD could still be Eq. (6). From human nature, humans may estimate the interpretation probability vector $P(y_H)$ with a limited number of classifications, and then get the interpretation result $y_H$ directly and immediately. Here $P(y_H)$ and $y_H$ can also be expressed as a one-hot vector. An example of recognizing a helmet image is shown in Fig. 4.

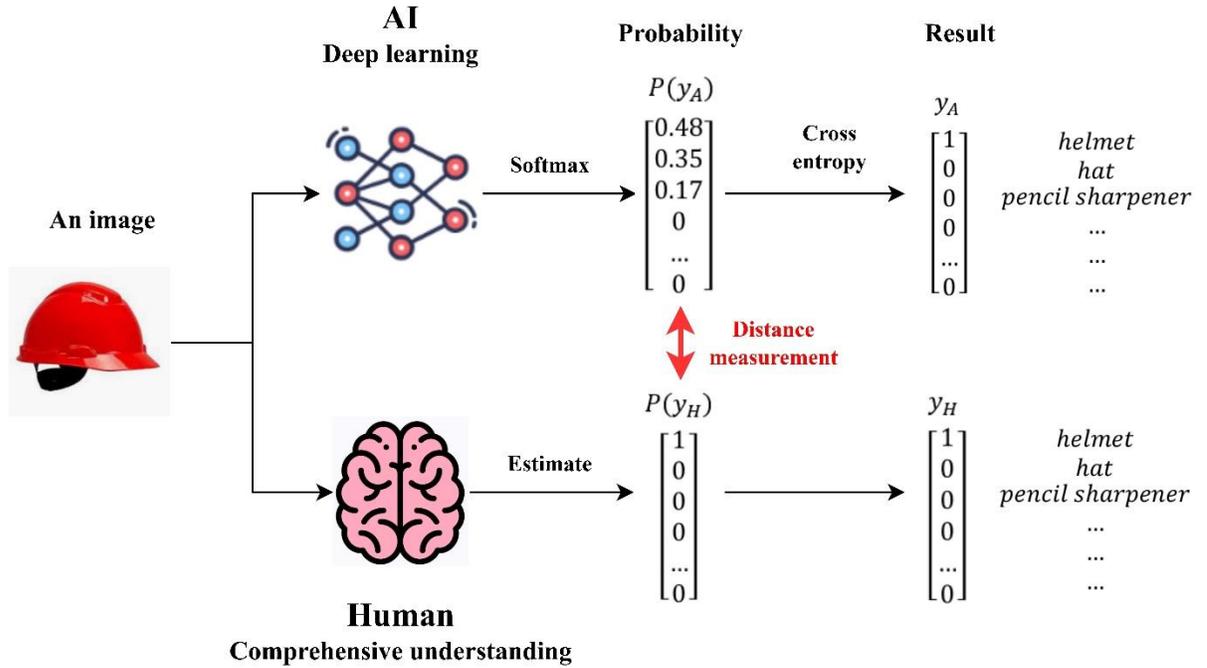

Fig. 4: An example of how humans and AI recognize an image (numbers are assumed).

Thus, interpretation conflict can be simplified as the difference between two one-hot vectors. Define the variable of interpretation difference (VID) as the difference in interpretation by different participants.

$$VID = y_A - y_H \tag{8}$$

When $VID = 0_{m \times 1}$, where $m$ is the number of classifications, there is no interpretation conflict; when $VID \neq 0_{m \times 1}$, there is an interpretation conflict.

Hence, a method to measure VID is by the distance from the AI probability vector to the human probability vector (Fig. 4), which can also be calculated with cross-entropy.

$$d_{VID} = cross\ entropy(P(y_A), P(y_H)) \tag{9}$$

### 3.3. Variable of action difference

Similarly, it can be defined that the variable of action difference (VAD) is the difference in control action by different participants.

$$VAD = U_A(t) - U_H(t) = \begin{bmatrix} u_{A1}(t) \\ u_{A2}(t) \\ ... \\ u_{Ap}(t) \end{bmatrix} - \begin{bmatrix} u_{H1}(t) \\ u_{H2}(t) \\ ... \\ u_{Hp}(t) \end{bmatrix} \qquad (10)$$

where $U_A(t)$ is the matrix of the controller action, $U_H(t)$ is the matrix of the human operator's control action, $u$ is the element of the action matrix, and the system has $p$ input variables.

Also, VAD can be measured as

$$d_{VAD} = \sqrt{\sum_{i=1}^{p}(u_{Ai}(t) - u_{Hi}(t))^2} \qquad (11)$$

### 4. Conflict risk assessment

Usually, observation conflict and interpretation conflict will cause human confusion or surprise, which will further affect human performance; consequently, action conflict may evolve into secondary disasters (Wen, Amin, et al., 2022).

The above derivation shows how distance can be used to measure conflict. When the distance is 0, the conflict probability is 0; when it reaches a certain value or more, the conflict probability is 1 (Fig. 5). Comparing any two time steps, the distance increases, indicating that the conflict possibility is increasing, and the conflict is diverging, or vice versa. Accordingly, suppose the probability follows the inverse function of a Beta distribution, and it can be expressed as

$$P = \begin{cases} BETA.INV(P, \alpha, \beta), 0 \leq d < d_{max} \\ \qquad\qquad 1, \qquad\quad d \geq d_{max} \end{cases} \qquad (12)$$

Where $P$ is the probability, $BETA.INV$ is to return the inverse of the beta cumulative probability density function, $\alpha$ and $\beta$ are the shaping parameters.

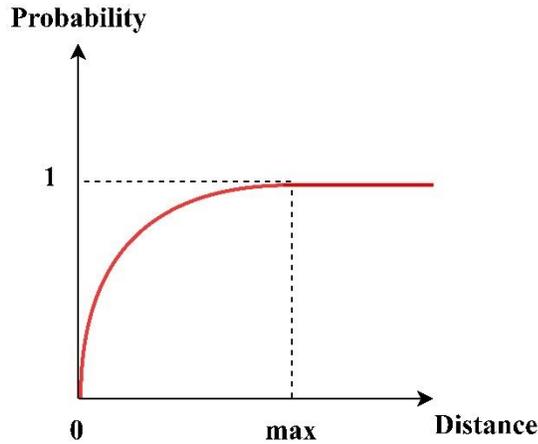

Fig. 5: The distribution of conflict probability.

Conflict severity is the consequence measure of a conflict. Naturally, when the distance increases, the conflict severity increases (Fig. 6). Therefore, the conflict severity could be proposed as

$$S = e^d - 1 \qquad (13)$$

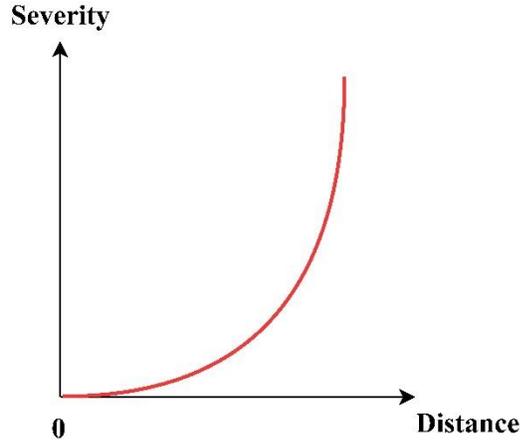

Fig. 6: The distribution of conflict severity.

Consequently, the conflict risk is

$$R = P \times S \qquad (14)$$

Usually, the risk can be graded and used to decide on strategy changes.

## 5. Discussion and conclusions

Human-AI conflict research's difficulty quantifies human observation, interpretation, and action. Currently, one proposal is to align them into the process variable, as all observation, interpretation, and action conflicts will finally reflect on the process variable. However, it is still debatable as this is not the nature of human observation, interpretation, and action. In addition, the human decision-making process is still a black box, and human actions are countless and unpredictable, which increases the difficulty of quantification, and it is difficult to reach a consensus between different scholars. Human-AI conflict is doomed to be a cross-research field involving sociology, engineering, computer science, and cognitive science.

Another research topic is whether the conflict can be resolved or resisted by advanced, robust, or AI control. Predictably, when the strength of fault, error, or attack is minimal, the controller may offset it, but when it reaches a particular strength, it is ineffective. The essence of conflict resolution would be to design conflict-tolerant control.

Of course, resolving conflicts from the human perspective is also possible. When the accuracy and reliability of AI are much higher than that of humans, whether AI has the power to exclude human interference involves the issue of priority. At present, AI has no value judgment, and the priority of AI should not be higher than that of humans. But it can also be seen that AI cannot resist human sabotage, which will still cause disaster. Perhaps, the conflict can be submitted to a third party for adjudication promptly and immediately. And this third party should still be a human being but not a party to the human-AI system.

In addition, this paper introduces some common cyberattacks. At the same time, there are more advanced network attacks, especially those targeting data, algorithms, and models, which are equivalent to attacking the eyes and brains of AI, such as data pollution, adversarial samples, backdoor attacks, etc. On the other hand, human sabotage is always unpredictable, and AI may not resist such conflicts. Therefore, more research is needed from the security perspective.

To be concluded, this study systematically introduces the topic of human-AI conflict from the definition, causes, measurement, and risk assessment. The results highlight that AI is the potential second decision-maker besides the human, and the human-AI conflict is a unique and emerging risk in industrial digitization.